# EASY-II: a system for modelling of n, d, p, γ and α activation and transmutation processes

Jean-Christophe Sublet [1*], James Eastwood [2], Guy Morgan [2], Arjan Koning [3] and Dimitri Rochman [3]

*[1] UK Atomic Energy Authority, Culham Science Centre, Abingdon OX14 3DB, United Kingdom, [2] Culham Electromagnetics Ltd, Culham Science Centre, OX14 3DB, United Kingdom, [3] Nuclear Research and Consultancy Group, NL-1755 ZG, Petten, The Netherlands.*
[*] Corresponding Author, E-mail:jean-christophe.sublet@ccfe.ac.uk

Abstract; EASY-II is designed as a functional replacement for the previous European Activation System, EASY-2010. It has extended nuclear data and new software, FISPACT-II, written in object-style Fortran to provide new capabilities for predictions of activation, transmutation, depletion and burnup. The new FISPACT-II code has allowed us to implement many more features in terms of energy range, up to GeV; incident particles: alpha, gamma, proton, deuteron and neutron; and neutron physics: self-shielding effects, temperature dependence, pathways analysis, sensitivity and error estimation using covariance data. These capabilities cover most application needs: nuclear fission and fusion, accelerator physics, isotope production, waste management and many more. In parallel, the maturity of modern general-purpose libraries such as TENDL-2012 encompassing thousands of target nuclides, the evolution of the ENDF format and the capabilities of the latest generation of processing codes PREPRO-2012, NJOY2012 and CALENDF-2010 have allowed the FISPACT-II code to be fed with more robust, complete and appropriate data: cross-sections with covariance, probability tables in the resonance ranges, kerma, dpa, gas and radionuclide production and 24 decay types. All such data for the five most important incident particles are placed in evaluated data files up to an incident energy of 200 MeV. The resulting code and data system, EASY-II, includes many new features and enhancements. It has been extensively tested, and also benefits from the feedback from wide-ranging validation and verification activities performed with its predecessor

## I. Introduction

FISPACT-II [1] is a completely new inventory code designed initially to be a functional replacement for FISPACT-2007. This new code is written in object-style Fortran 95 and has extended physical models, a wider range of irradiation options and improved numerical algorithms compared to the old code. Users familiar with the old code will be able in most cases to use the new code with their existing control input files. Some noteworthy new keywords have been added to deal with the new capabilities, and some of the old keywords have become obsolete.

The major change introduced in this first release of FISPACT-II was the addition of the reading and processing of alternative ENDF-format library data sets for different incident particles. This has caused a major overhaul of the data input parts of the software and a huge expansion of the number of nuclides and reactions that can be treated. Sensitivity and error prediction capabilities have been extended, and better fission yield data and cross-section data in more energy groups up to higher energies can now be used. The present version can also handle more irradiating projectiles (α, γ, n, p, d) and provides additional diagnostic outputs (prompt and delayed kerma, dpa and gas appm rates) if the ENDF-format library contains the required input data. The new code can also connect to any version of earlier EAF-formatted libraries [2].

The new inventory code when associated with a set of nuclear data libraries (ENDF/Bs, EAFs or TENDLs [3]), plus decay, biological, clearance and transport index libraries, forms the European Activation System EASY-II.

## II. The models

The FISPACT-II code follows the evolution of the inventory of nuclides in a target material that is irradiated by a time-dependent projectile flux $\varphi$, where the projectiles may be neutrons, protons, deuterons, α-particles or γ-rays. The material is homogeneous, infinite and infinitely dilute and the description of the evolution of the nuclide numbers is reduced to the set of stiff ODEs Eq. (1) for $N_i$ the number of atoms of nuclide $i$ [4]. The key characteristics of the system of inventory equations are that they are linear, stiff and sparse.

$$\frac{dN_i}{dt} = -N_i(\lambda_i + \sigma_i\varphi) + \sum_{j\neq i} N_j(\lambda_{ij} + \sigma_{ij}\varphi) \qquad (1)$$

Here $\lambda_i$ and $\sigma_i$ are respectively the total decay constant and cross-section for reactions on nuclide $i$, $\sigma_{ij}$ is the cross-section for reactions on nuclide $j$ producing nuclide $i$, and for fission it is given by the product of the fission cross-section and the fission yield fractions. $\lambda_{ij}$ is the constant for the decay of nuclide $j$ to nuclide $i$.

The stiffness of the system of equations limits the choice of numerical methods. The code uses the Livermore solver for ordinary differential equations LSODES [5] to solve the stiff ODE set. LSODES implements the Backward Differentiation Formula (BDF also known as Gear's method) and uses the Yale sparse matrix package to handle the Jacobian matrices. An Adams method (predictor-corrector) is used in non-stiff cases. This numerical solver compares advantageously with the previous EXTRA ODE solver, written in 1976, and used in all versions of FISPIN and FISPACT. FISPACT-II has a wrapper ODE module around LSODES that automatically sets storage, parameters and dynamic memory allocation for that solver, improving portability and reducing the need for user input.

Note that FISPACT-II differs from FISPACT-2007 in that it does not employ the equilibrium approximation for short-lived nuclides, and includes actinides self-consistently in the rate equations (Eq.(1)) rather than as a source term. The new code has been shown to be able to handle short (1ns) time interval and high flux cases that caused problems for older codes.

## III. Nuclear data libraries

FISPACT-II requires connection to several data libraries before it can be used to calculate inventories. The following libraries are required: cross-section data for projectile-induced reactions, uncertainty data for neutron-induced reactions, decay data, fission yields, biological hazard, legal transport, clearance and gamma absorption data. It is a user choice to select from the library versions. Any libraries in the correct ENDF-6 format could be used. The development of FISPACT-II over the last few years has run in parallel with the development of the TALYS-based [6] Evaluated Nuclear Data Library (TENDL) project and those latest fully fledged European libraries are now the recommended source of nuclear data [3]. The TENDL libraries have made possible new predictive capabilities.

**1. Cross-sections**

The TENDL-2012 library [3] is the current recommended evaluated data source for use in any type of nuclear technology applications. The principal advances of this new library are in the unique target coverage, 2424 nuclides; the upper energy range, 200 MeV; variance and covariance information for all nuclides; and the extension to cover all important projectiles: neutron, proton, deuteron, alpha and gamma, and last but not least the proven capacity of this type of library to transfer regularly the feedbacks of extensive validation, verification and benchmark activities from one release to the next.

The TENDL libraries processed in line with the ENDF format framework brings the new activation prediction capabilities. The pointwise data are available from $10^{-5}$eV to 200 MeV, including resonance ranges. Elastic cross-section data are available for the self-shielding corrections (see Sec IV.1). Gas production, dpa and kerma cross-sections are used to give gas appm, displacement per atom and kerma diagnostics in activation calculations. Gamma and alpha reaction cross-sections introduce new classes of calculations. TENDL-2012 is the fifth generation of such a library and as such has benefited from the previous releases and from the EAF-2010 V&V processes [7].

The cross-section data are provided in two universal fine group structures: a CCFE (709 groups) scheme for the neutron-induced cross-sections and a CCFE (162 groups) scheme for the non-resonant p, d, $\alpha$ and $\gamma$–induced cross-sections. The data format used is fully compliant with the ENDF-6 manual specification handled on an isotopic basis and so allows many existing utility codes to further manipulate, visualise or check any aspects of the pre-processed files. The data files are produced using a complex but robust, complementary sequence of modules of the processing codes NJOY99.396, PREPRO-2012 and CALENDF-2010 [7].

**2. Fission yields**

The fission yield data need to be provided for each actinide and incident particle. Only 19 of the many nuclides that undergo fission have any fission yield data in JEFF-3.1.1 and these cover only a reduced energy range. For the remainder the UKFY4.1 library [8] then further extends the target and energy range before a neighbouring fission yield is used. This UKFY4.1 library using Wahl's systematics is also used for all other particle-induced fission yields.

**3. Decay data**

In addition to cross-sections the other basic quantities required by an inventory code are information on the decay properties (such as half-life) of all the nuclides considered. FISPACT-II is able to read the data directly in ENDF-6 format; it requires no pre-processing to be done. The now well-verified and validated EAF_dec_2010 library based primarily on the JEFF-3.1.1 and JEF-2.2 radioactive decay data libraries, with additional data from the latest UK evaluations, UKPADD-6.12, contain 2233 nuclides. However, to handle the extension in incident particle type, energy range and number of targets many more decay data are needed. A new 3873-nuclide decay library dec_2012 has been assembled from EAF_dec_2010 complemented with all of JEFF-3.1.1 and a handful of ENDF/B-VII.1 decay files.

**4. Radiological data**

The radiological data for the increased number of nuclides present in the TENDL-2012 data are computed in the same manner as described for the EAF data. The new hazards, clearance and transport data are respectively for 3647, 3873 and 3872 nuclides, compared to 2006, 2233 and 2233 for the EAF data.

**5. Variance-covariance**

Above the upper energy of the resolved resonance range, for each of the 2424 nuclides a Monte Carlo method in which the covariance data come from uncertainties of the nuclear model calculations is used. In the TENDL-2012 library, all information on cross-section covariance is stored in the MF=33 or 40 formats, starting at the end of the resonance range up to 200 MeV. Short-range, self-scaling variance components are also specified for each MT type. The data format used to store the variance-covariance information has been made fully compliant with the ENDF-6 format description and the files are read directly by FISPACT-II without any further intermediate processing. Variance and covariance data are used by FISPACT-II to create uncertainty predictions and sensitivity analyses.

## IV. Advanced physics

The original version of FISPACT-II was engineered to be a functional replacement for the earlier FISPACT-2007code, but it has now been extended to include major new capabilities and enhancements.

### 1. Self shielding of resonant channels.

The CALENDF-2010 [9] nuclear data processing system is used to convert the evaluation defining the cross-sections in ENDF-6 format (i.e., the resonance parameters, both resolved and unresolved) into forms useful for applications. Those forms used to describe neutron cross-section fluctuations correspond to "cross-section probability tables", based on Gauss quadrature and effective cross-sections. FISPACT-II iteratively solves for the dilution cross-section (which depends on mixture fractions and total shielded cross-section) and the shielded cross-section for nuclides in the mixture (which depends on dilution cross-section and probability table data) [1,9].

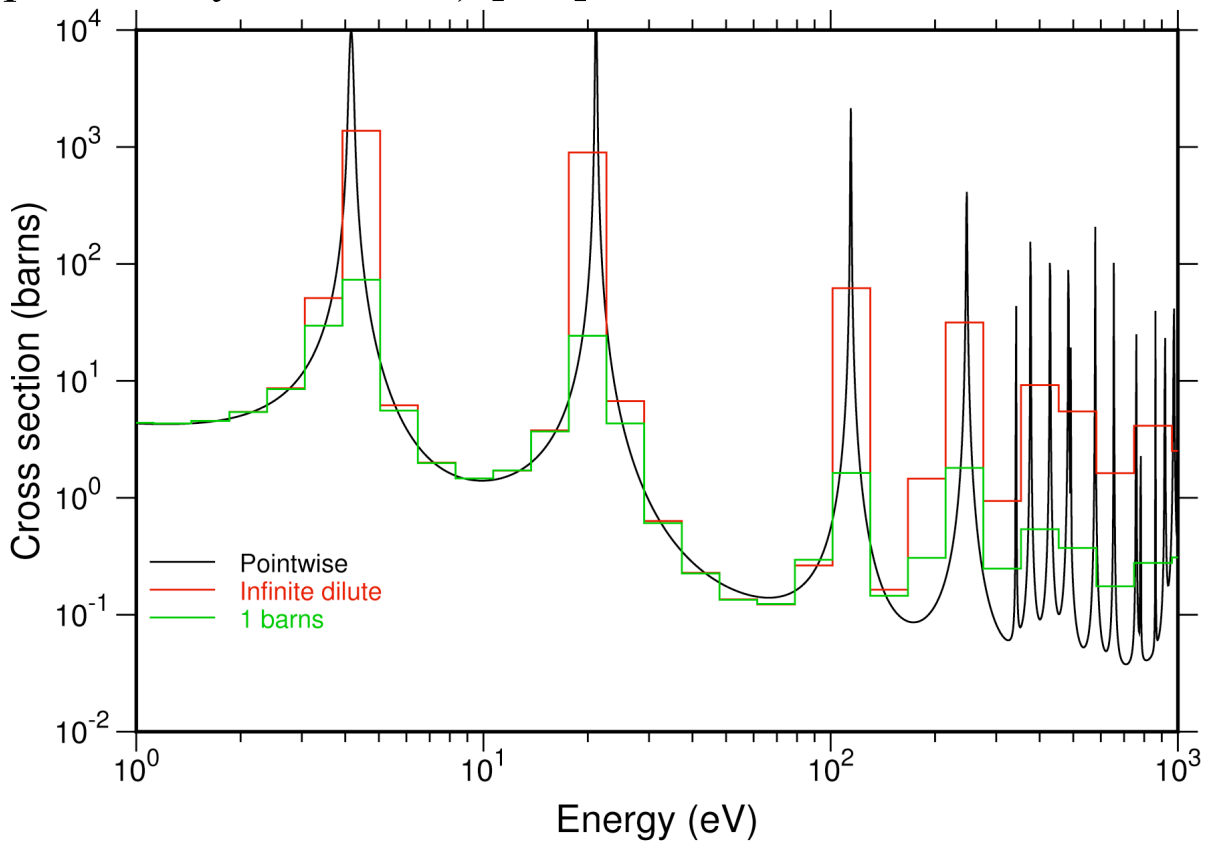

**Figure 1: Self-shielding effect for different dilutions, which show that changes to effective groupwise cross-section may be large**

CALENDF-2010 provides probability tables in the energy range from 0.1 eV up to the end of the resolved or the unresolved resonance range. Probability table data in 709 energy group format are provided for 2314 nuclides of the TENDL-2012 library. These data are used to model dilution self-shielding effects from channel, isotopic or elemental interferences (c.f. Figure 1). Doppler broadening effects are also included and the tables are given at three temperatures: 293.6, 600 and 900 degree Kelvin.

The dilution cross-sections computed using the CALENDF data are applied either as scaling factors to the library cross-section data or as replacements over the energy ranges for which the probability table data are available [1]. This ability to self-shield, in much the same manner as is done in deterministic transport codes and in Monte Carlo codes for the unresolved resonance range (URR) depicted is believed to be unique amongst inventory codes.

### 2. Pathways

The reaction network may be described either by the rate equations or as the sum of paths and loops, which we refer to as pathways. The inventory of a given nuclide computed using the rate equations can equivalently be found by a linear superposition of contributions of flows along the pathways to that nuclide. Pathways analysis is used in identifying significant nuclides and reactions, and in performing sensitivity and uncertainty analyses [10].

Pathways analysis uses directed graph algorithms implemented using breadth-first tree searches with pruning for finding routes from a parent to chosen descendants, and for the assembly and solution of a subset of the rate equations for nuclides on a pathway to get the flow along that pathway. Pathways analyses may be performed for single and multiple step irradiation scenarios, and where the cross-sections are time dependent.

### 3. Uncertainty estimates

Pathways analysis identifies the pathways from the initial inventory nuclides to the (target) dominant nuclides at the end of the irradiation phase, and provides the number of atoms of each nuclide produced by reactions and decays along each pathway. These, together with uncertainties derived from the covariances in the reaction cross-sections and decay half-lives associated with the edges of the pathways are used in FISPACT-II to provide estimates of the uncertainties [10]. The uncertainties are then computed for all significant radiological quantities, e.g., number density, activity, decay heat, dose rate, inhalation or ingestion hazards.

More accurate uncertainty estimates that may also include covariance between different reaction cross-sections can be undertaken by combining pathways analysis with Monte-Carlo sensitivity calculations.

### 4. Thin and thick target yields

A second method of accounting for self-shielding in thick targets with a variety of geometries has been provided. The development is based on the experimental and theoretical work of Baumann [11] described as a "universal sigmoid curve" model of self-shielding by Martinho, Gonçales and Salgado [12,13]. The Martinho [12] model initially described the effect of a single resonance peak in a pure target consisting of a single nuclide. The self-shielding factor $G_{res}$ is approximated as a simple function of a single dimensionless length parameter that depends on the physical size and shape of the target as well as the peak cross-section at the resonance and the resonance widths for elastic scattering and radiative capture. The final form of the model [13] accommodates a group of isolated resonances of a pure target, and the target geometry could be a foil, wire, sphere or cylinder of finite height.

In more detail, the initial form of the model that accounts for the effect of a single resonance in a pure target containing a single nuclide defines a dimensionless parameter

$$z = \sum_{tot}(E_{res})\, y \sqrt{\frac{\Gamma_\gamma}{\Gamma}} \qquad (2)$$

that depends on the physical length y, the macroscopic cross-section at the energy $E_{res}$ of the resonance peak, the resonance width $\Gamma_\Upsilon$ for radiative capture and the total resonance width Γ. Then the self-shielding factor can be expressed as

$$G_{res}(z) = \frac{A_1 - A_2}{1 + \left( z/z_0 \right)^p} + A_2 \qquad (3)$$

Where the parameters defining the "universal sigmoid curve" allows the fitting of the experimental data regardless of the geometry and nuclides.

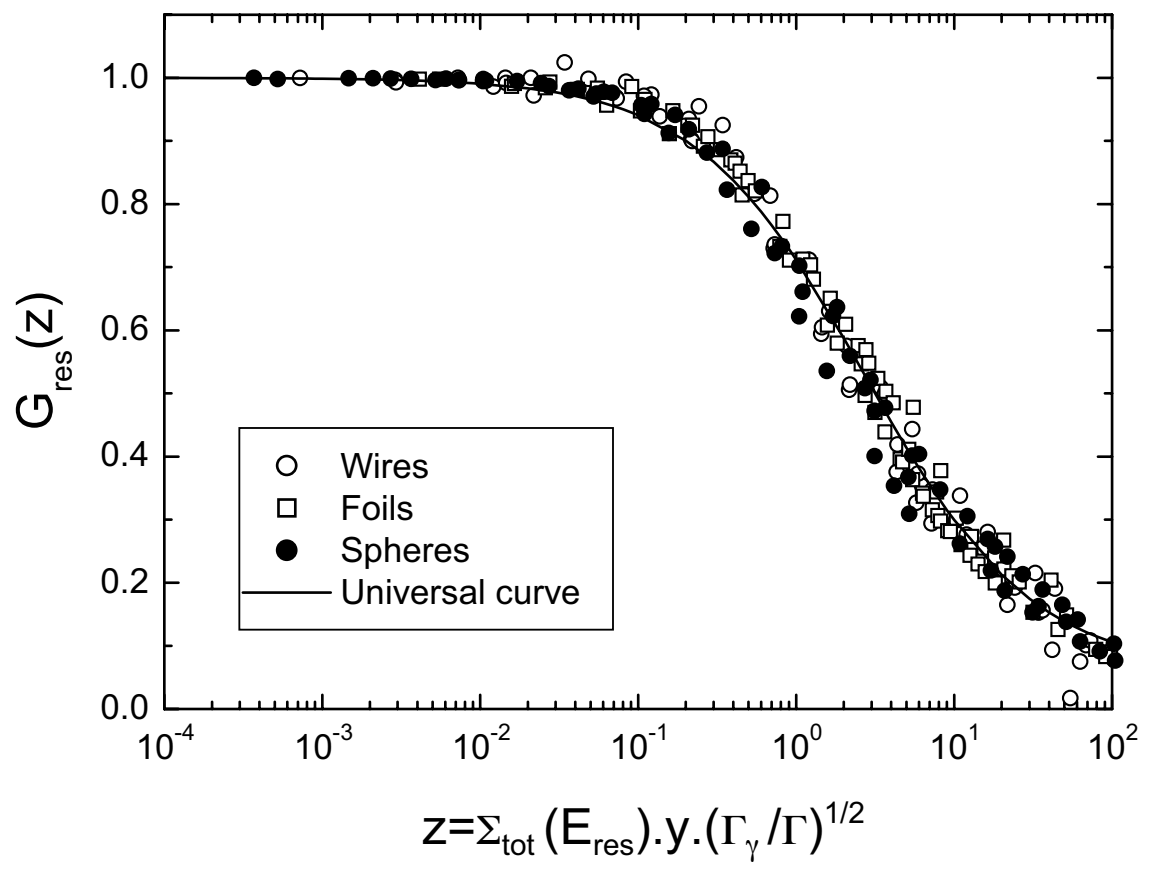

**Figure 2: Universal curve of resonance self-shielding factor as a function of z and target shape [12]**

This model has been generalised further in two ways to make it suitable for applications and to the mixture of nuclides required for a FISPACT-II calculation.

First, the average self-shielding factor is computed from the resonance parameters given in the resolved resonance range defined in the ENDF file 2 data for a subset of the nuclides specified with keywords. Note that TENDL-2012 uses a unique approach to create parameters where none were known previously for resolved statistical resonances for a large number of nuclides living longer than one second [15]. The cross-section at a resonance peak is not supplied in the groupwise ENDF data. The simple expression provided by Fröhner [15] is used to calculate this information.

Secondly, $G_{res}$ is made energy dependent by taking averages separately for each energy bin used for the group-wise cross-sections, including only those resonances with peaks in the relevant energy bin. Then this array of energy-dependent self-shielding factors is applied to each energy-dependent cross-section before the cross-section collapse.

The principle underlying this model of self-shielding is that the resonances perturb the spectrum of the applied neutron flux. Consequently, the self-shielding factors should modify the cross-sections for all reactions. However, the effect of self shielding varies from reaction to reaction because of the differing energy dependencies of the cross-sections.

## V. Verification and Validation

Verification and Validation (V&V) is a critical, yet often overlooked, part of scientific computer code development. Verification is the process of determining whether or not the products of a given phase in the software life cycle fulfil a set of established requirements. This implies an on-going process of unit testing to ensure that the algorithms, which are implemented in the code, solve the correct equations in order to calculate the required quantities. In contrast, Validation is the stage in the software life cycle at the end of the development process where software is evaluated to ensure that it complies with the requirements. This is a more comprehensive effort, which is intended to test the code in aggregate to ensure that the code is obtaining the correct results for the required quantities.

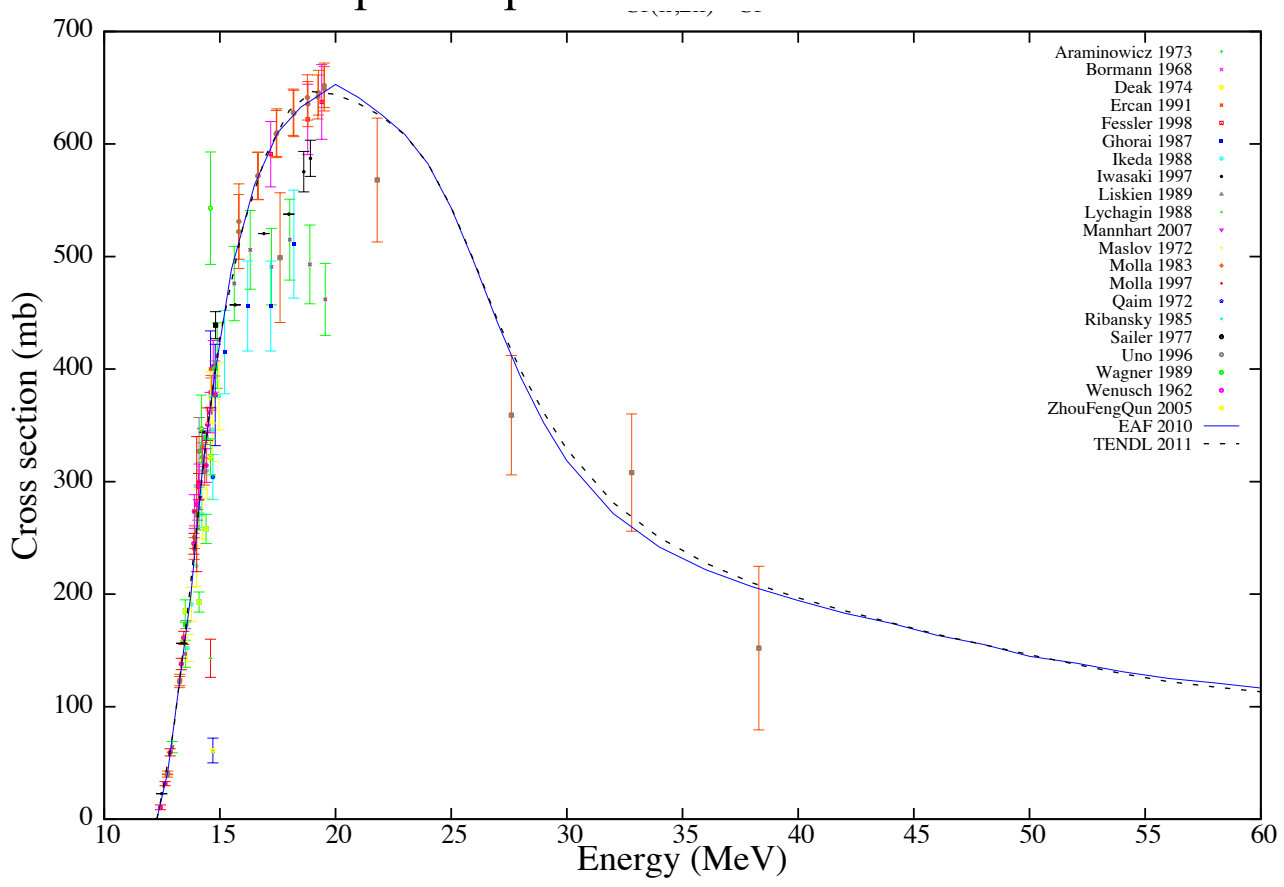

**Figure 3: TENDL-2012, EAF-2010 Cr-52(n,2n) cross section against EXFOR experimental data**

Careful software lifecycle management under configuration control has been used for the code, unit and integration tests and verification tests. EASY-II is distributed with over 400 input/output regression tests that preserve and extend the verification-validation heritage of EASY-2010 and its predecessors [7]. Validation usually relate to comparison with differential and/or integral experimental information. Library dependent pointwise cross-section, excitation function data can be plotted against the differential experimental information contained in the EXFOR data base. The data base contains the measured cross-section value at a given neutron energy usually with its associated uncertainty.

The EXFOR collection reached a milestone in 2013 in accumulating information for more than 20,000 different experimental works, many including uncertainty ranges.

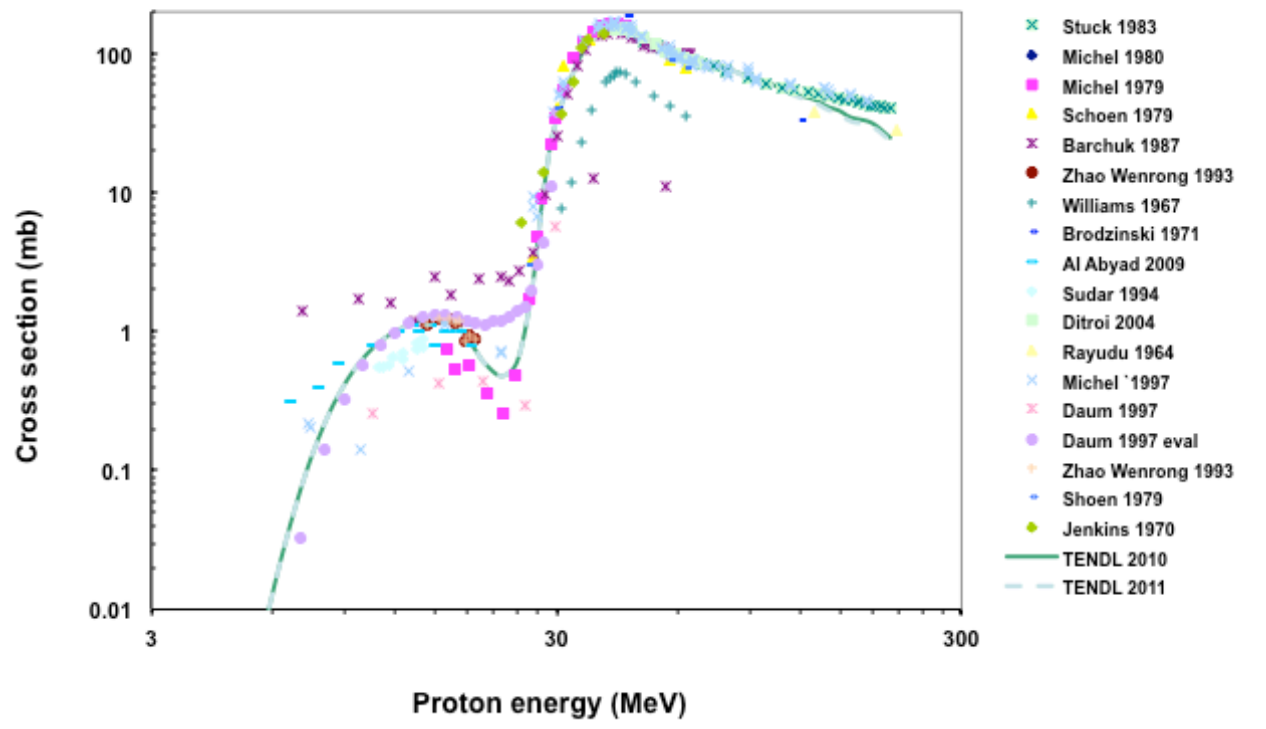

**Figure 4: TENDL Fe-nat(p,x)Mn-54 productions channels against EXFOR experimental data**

The EXFOR collection reached a milestone in 2013, in accumulating information for more than 20,000 different experimental works, many including uncertainty band.

The systematics of neutron cross-section can also be use to statically analysed the excitation function shape.

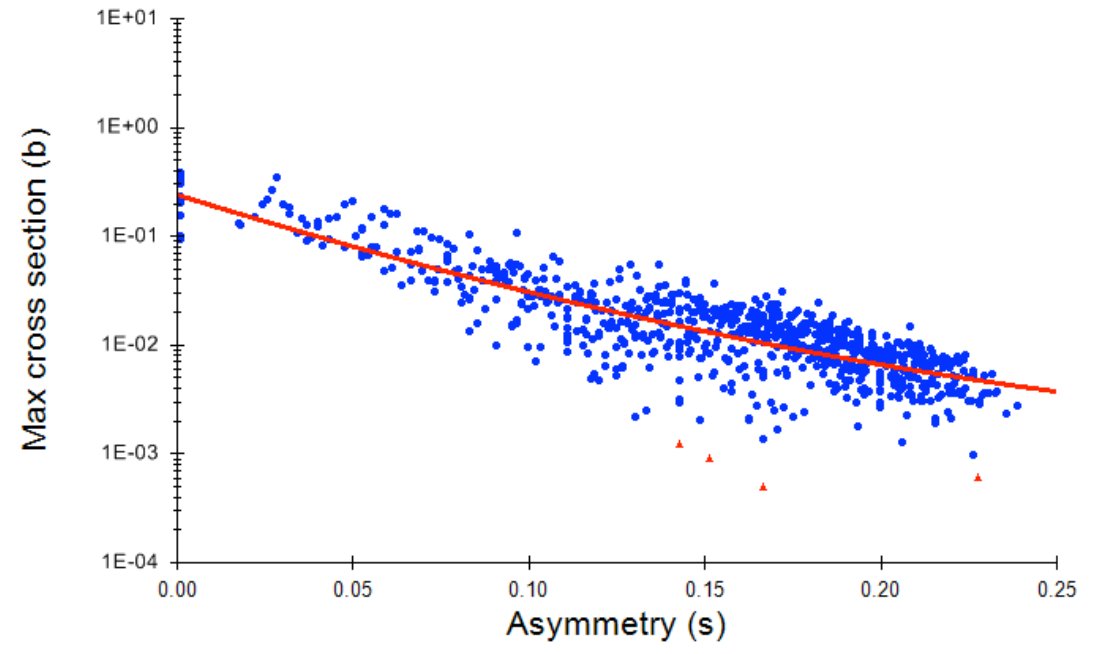

**Figure 5: TENDL-2012 (n,α) maximum cross section against asymmetry parameter**

Reaction rates, effective cross-section validation, integral validation is also very useful. When simulated integral values are compared against experimental measurements. For that the JAEA FNS assembly where 14 MeV neutrons are generated by a 2 mA deuteron beam impinging on a stationary tritium bearing titanium target has proven most beneficial [16].

More than seventy different material samples have been irradiated in sequence to have their decay heat measured in a whole energy absorption spectrometer. The value of those integral results reside mainly in the well characterized and stable neutron spectra at the target position, but also to the time scale of the measurements from a few seconds after irradiation up to 400 days. All experimental results have been compared with values derived from the EASY-II simulation. For both calculated and experimental values uncertainty estimates are also provided.

Figure 6 demonstrates the validation results for two samples of Yttrium oxide irradiated for 5 minutes and 7 hours at JAEA FNS. Decay heat was measured for both samples at many cooling time ranging from 36 s to 400 days. The short term decay heat is well predicted, but the code prediction seems to underestimate the long term one by around 20%.

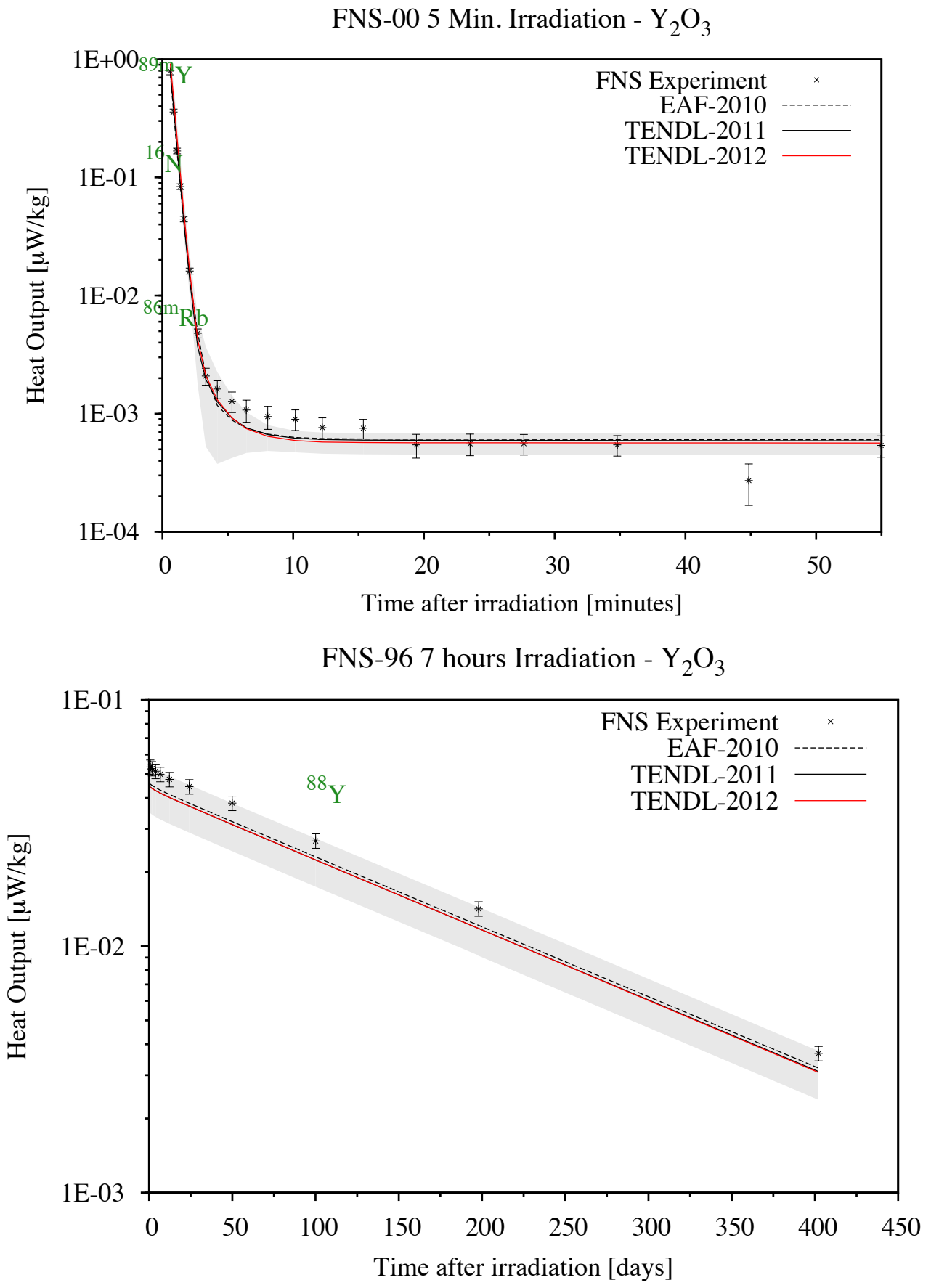

**Figure 6: Yttrium oxide decay heat comparison, TENDL derived uncertainty as gray area.**

Further V&V processes are being actively deployed and demonstrated in support of EASY-II and the associated TENDL libraries.

## VI. EASY-II capabilities

- Large number of targets: 2434 from $H^1$ to $Ds^{281}$
- Broad energy range: $10^{-5}$eV to 200 MeV
- Five incident particles: α, γ, d, p, n
- Variance-covariance, uncertainty
- Pathways analysis, dominant nuclide
- Number density, activity, decay heat, dose rate, inhalation, ingestion, clearance, transport indices
- Self-shielding: channel, isotopic, elemental
- Sensitivity analysis (Monte Carlo)
- DPA, Kerma, gas and radionuclide production
- Thin, thick target yields

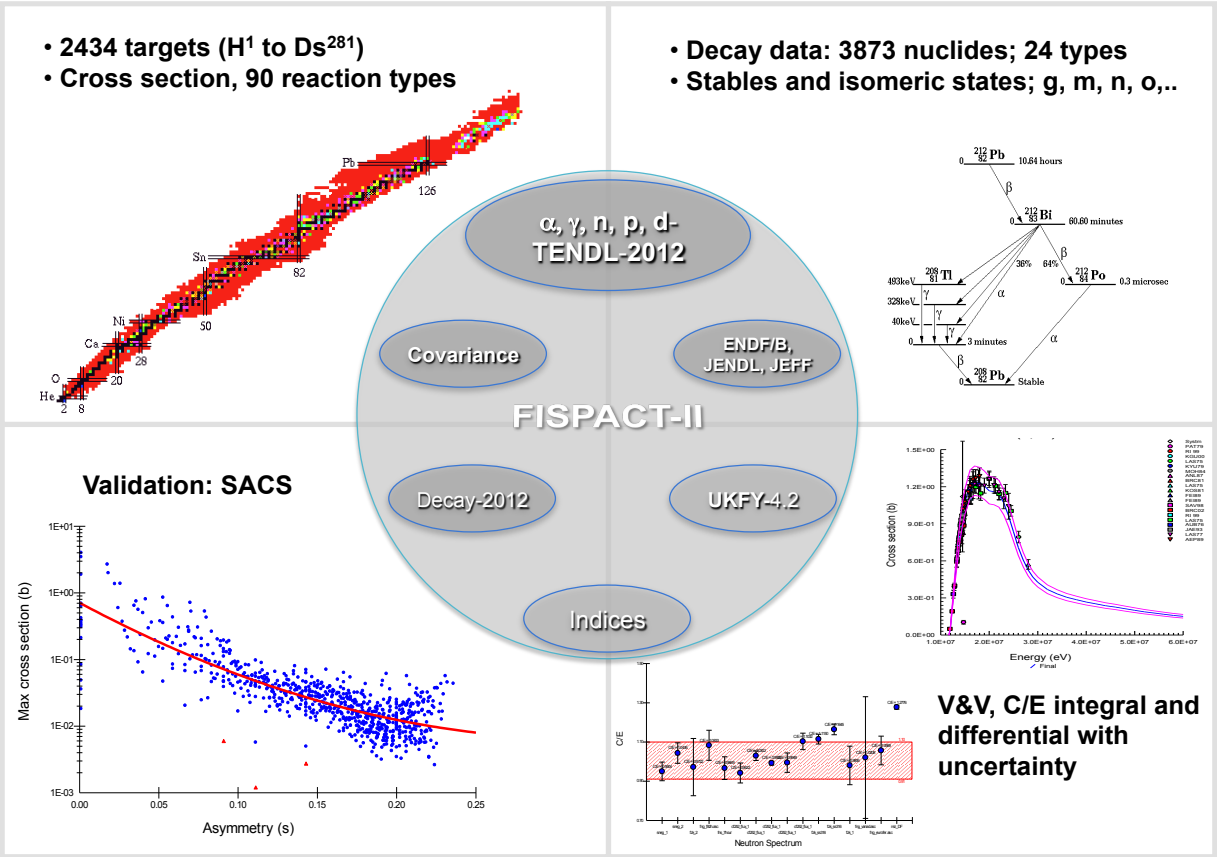

**Figure 7: EASY-II: FISPACT-II & TENDL's libraries**

## VII. Conclusions

EASY-II is a new and versatile multi-particle inventory code and nuclear data package aimed at satisfying all burnup, buildup, activation-transmutation, decay requirements for facilities in support of any nuclear technology: stockpile and fuel cycle stewardship, materials characterization and life cycle management. It has been developed and tested for: magnetic and inertial confinement fusion, fission Gen II, III, IV plant generations. It is also applicable to high energy and accelerator physics; medical applications, isotope production, earth exploration and astrophysics.

## Acknowledgements

This work was funded by the RCUK Energy Programme under grant EP/I501045. To obtain further information on the data and models underlying this paper please contact PublicationsManager@ccfe.ac.uk.